\documentclass[a4paper]{article}
\usepackage[english]{babel}
\usepackage[utf8]{inputenc}

\usepackage[nottoc]{tocbibind}
\usepackage{authblk}
\usepackage{amsmath}
\usepackage{amsthm}

\newtheorem{proposition}{\indent Proposition}

\title{APPLICATION OF DIAGNOSTIC TEST METHODS TO THE CLASSIFICATION OF TIME SERIES WITH DISCRETE VALUES}

\author[1]{Artyom Gevorgyan}
\affil[1]{School of Electronic Information and Electrical Engineering, Shanghai Jiao Tong University \protect\\ \texttt{ar.gevorgyan2002@sjtu.edu.cn}}

\author[2]{Albert Gevorgyan}
\affil[2]{Department of Mathematics, Imperial College London \protect\\ \texttt{a.gevorgyan20@imperial.ac.uk}}

\begin{document}

\maketitle

\begin{abstract}
Discrete-value time series are sequences of measurements where each measurement is a discrete (categorical or integer) value. These time series are widely used in various fields, and their classification and clustering are essential for data analysis. This article presents the possibility of applying diagnostic test methods to such time series and estimates the probability of finding ``matching tests''.
\end{abstract}

\section*{Introduction}
Discrete value time series are sequences where each measurement has a discrete value. These series differ from time series with continuous values, where each measurement can be any number from a continuous range.

Time series with discrete values can be used to model and analyze various types of data, including categorical data and integer data. In the case of categorical data, each time series value is one of a limited set of categories or classes. For example, a time series may contain data on the health status of a patient (normal, sick, rehabilitated), the quality of a product (good, defective), or the change in asset prices in financial markets (price rose, fell, or remained unchanged).  In the case of integer data, each dimension is an integer. Examples are daily sales of goods, the number of cab or ambulance calls during the day, or the number of visits to a website in each hour. Another example of a time series with discrete values would be the number of transactions in a financial asset during an hour.

It is important to note that analyzing time series with discrete values may differ from analyzing time series with continuous values. In particular, such series may require special methods and models that take into account the discreteness of the data.

This paper proposes a method of applying diagnostic tests for discrete time series. In frequency, it is proposed to consider a matrix in which each row is a time series.

\section*{Literature Review}
The diagnostic test method is commonly utilized in pattern recognition and classification tasks. In his work, Liu \cite{liu_motoda} presents a comprehensive overview of different techniques for selecting features and variables. The book also includes a section on diagnostic tests and their application in pattern recognition and classification tasks. Dyukov and Peskov \cite{kiam_4481}, on the other hand, focus on complex model-building problems that involve numerous relationships. The authors offer precedent-based methods as a solution to such problems in pattern recognition. To classify and cluster time series, it is common in the literature to reference nearest neighbor (k-NN) techniques, which can be modified for use with discrete-valued time series data. In order to measure the similarity between time series, specialized metrics, such as Hamming distance \cite{nnb}, are employed.

Another commonly employed technique for analyzing time series with discrete data involves using feature-based methods. In this approach, important features are extracted from time series values which take on discrete values. These extracted features might include statistical characteristics like category or integer distributions and temporal characteristics such as the frequency and duration of certain events \cite{Fulcher_2014}.

In recent years, there has been a rise in the adoption of machine learning techniques. Time series with discrete values can also be analyzed using convolutional neural networks (CNN) and recurrent neural networks (RNN). These networks can analyze the sequence of data and identify intricate dependencies \cite{10.1007/978-3-319-52962-2_14}.

The practical use of pattern recognition and classification techniques in analyzing discrete time series data is extensive. In medicine, such methods aid in diagnosing diseases through test results and patient symptom analysis \cite{cdec8a262ac44f038fdd72b16a758736}.

The methods discussed are frequently employed in predicting asset price fluctuations in the stock market. Explanatory variables with integer and categorical values are commonly found in such studies. Sonkavde et al. offer a thorough and inclusive analysis of these methods and their outcomes across various instruments and securities markets \cite{ijfs11030094}.

\section*{Model Description}

Discrete math methods are frequently used in pattern recognition issues. One of the most renowned methods is founded on testing \cite{yankovskaya2016accelerated, zhuravlev2006, zhuravlev78prob33}. Consider a matrix with dimensions $n\times m$, where the rows denote ``objects'' and the columns denote ``features''. Every element $a_{ij}$ within the matrix, where $i=1,2,..,n$; $j=1,2,..,m$ denotes the assessment of the $i$-th object by attribute $j$. 

In this paper, we propose a method for considering time series as the rows of matrix $A^{'}$, where the columns represent time intervals. For clarity, let us examine a practical example. Suppose we examine the daily time series of several assets in the stock market. If we track the price changes of five assets over a month, we can create a matrix of dimensions $5\times 31$.
In this case, the definition of the new matrix $A^{'}$ is as follows: 

\[
a'_{ij} =
\begin{cases}
  1 & \text{if } a_{i+1,j} \geq a_{ij}, \\
  0 & \text{if } a_{i+1,j} < a_{ij}.
\end{cases}
\]

For each $i=1,2,..,n$.

The matrix is a $5\times 30$ dimensional array where all elements consist of either $0$ or $1$. A value of $1$ signifies a positive return for asset $i$ at moment $j$, whereas $0$ indicates the opposite outcome.

Diagnostic test methods can be used to obtain homogeneous clusters of the stock 
market, and to identify significant days contributing to market segmentation. 

These findings have numerous applications, from forecasting price changes to constructing optimal asset portfolios. The proposed method does not adhere to the typical prerequisites of classical time series models, including the necessity for the series to be stationary.

The findings from \cite{abo2018} demonstrate the ability to calculate the probabilities of attaining ``matching'' tests, thus allowing for the estimation of diagnostic test parameters for time series classification.

\section*{Results}

A matrix test is a submatrix created by removing some columns from a matrix in such a way that any two rows in the resulting matrix are different. A test is said to be a ``dead-end'' test if none of its submatrices is a test. In pattern recognition problems, binary matrices are typically used, which simplifies the process of identifying dead-end tests. In \cite{abo2018}, the concept of a dead-end test is extended to matrices, with each element capable of taking $k$ distinct values. The probabilities of discovering dead-end tests are calculated based on the matrix dimensions and the value of $k$, demonstrating that the scope for discovering such tests is limited.

Two rows, $i_1$ and $i_2$, are regarded as matching if all elements in row $i_1$ are identical to those in row $i_2$, where $i_1,i_2 = 1,2,..,n$.

Let $P$ be the probability of a match between any two lines $i_1$ and $i_2$ in the matrix $A^{'}$.

\begin{proposition} The probability of matching any two rows of matrix A is determined by the following expression:

\begin{equation*}
    1-\frac{k^{m}(k^{m} - 1)...(k^{m} - n + 1)}{k^{mn}}
\end{equation*}

\end{proposition}

Let for each row of matrix $j$ a different value $k_1,k_2,...,k_m$ is defined. Let us denote such a matrix by $B$.

\begin{proposition} The probability of matching any two rows of matrix B is determined by the following expression:

\begin{equation*}
    1 - \frac{{k_1}{k_2}...{k_m}({k_1}{k_2}...{k_m} - 1)...({k_1}{k_2}...{k_m} - n + 1)}{({k_1}{k_2}...{k_m})^n}
\end{equation*}
    
\end{proposition}

Let's call a submatrix of a matrix of dimension $l\times m$, where $l < m$ and all the rows match, a ``matching'' submatrix. For a dead-end test of dimension $l + 1$ to exist, a matching matrix of dimension $l$ must exist.

\begin{proposition} The probability of finding a "matching test" of dimension $l$ is equal to:

\[
\sum_{\substack{|S| \geq l }} \frac{\prod_{i \in S} (k_i^{n-1} - 1)}{(k_1k_2 \cdots k_m)^{n-1}}
\]  

\end{proposition}

\section*{Conclusion}
Based on these statements, the following conclusions can be drawn:
\begin{itemize}
    \item The probability of identification of a dead-end test of order $log_{k} n$ is substantial, and hence its detection is possible. The greater the length of the dead-end test, the higher the amount of information it contains, leading to an increased possibility of discovering patterns. This, in turn, enhances the efficiency of discretization.
    \item The probability of finding a dead-end test of length $l$ decreases more slowly with increasing number of time series compared to increasing length of time series and value of $k$, in the case where $m > 2$.
    \item When analyzing the time series of asset prices in stock portfolio management, it is advisable to analyze short time series with a small discretization level for a sufficiently large number of assets, based on the obtained results. The proposed method shows better performance for short time series compared to classical classification methods.
\end{itemize}

\bibliographystyle{acm}
\bibliography{main}

\end{document}